\documentclass[journal=jpclcd, manuscript=letter, layout=twocolumn
]{achemso}
\usepackage{amsfonts,amsmath,bm,multirow}
\usepackage[OT4]{fontenc}

\usepackage{etoolbox}
\usepackage{soul}

\usepackage{xcolor}
\usepackage{graphicx}
\usepackage{physics}
\usepackage{hyperref}
\usepackage{cleveref}
\usepackage{dcolumn}
\usepackage{subfigure}
\usepackage{float}

\makeatletter
\def\@email#1#2{
 \endgroup
 \patchcmd{\titleblock@produce}
  {\frontmatter@RRAPformat}
  {\frontmatter@RRAPformat{\produce@RRAP{*#1\href{mailto:#2}{#2}}}\frontmatter@RRAPformat}
  {}{}
}
\makeatother

    \title{Hydrostatic Pressure Driven Band Gap Tuning and Self-Trapped Exciton Formation in (4FPEA)$_2$SnBr$_{4}$ Halide Perovskite}

\author{Rafał Bartoszewicz}
\affiliation{Department of Semiconductor Materials Engineering, Wrocław  University of Science and Technology, Wybrzeże Wyspiańskiego 27, 50-370 Wrocław, Poland}
\email{rafal.bartoszewicz@pwr.edu.pl}
\author{Jakub Ziembicki}
\affiliation{Department of Semiconductor Materials Engineering, Wrocław  University of Science and Technology, Wybrzeże Wyspiańskiego 27, 50-370 Wrocław, Poland}
\author{Ewelina Zdanowicz}
\affiliation{Department of Semiconductor Materials Engineering, Wrocław  University of Science and Technology, Wybrzeże Wyspiańskiego 27, 50-370 Wrocław, Poland}
\author{Artur P. Herman}
\affiliation{Department of Semiconductor Materials Engineering, Wrocław  University of Science and Technology, Wybrzeże Wyspiańskiego 27, 50-370 Wrocław, Poland}
\author{Jesús Sánchez-Diaz}
\affiliation{Institute of Advanced Materials (INAM), Universitat Jaume I. Av. de Vicent Sos Baynat, Castellón de la Plana, 12006 Spain}
\author{Samrat Das Adhikari}
\affiliation{Institute of Advanced Materials (INAM), Universitat Jaume I. Av. de Vicent Sos Baynat, Castellón de la Plana, 12006 Spain}
\alsoaffiliation{Institute of Physical Chemistry, Polish Academy of Sciences, Warsaw 01-224, Poland}
\author{Iván Mora-Seró}
\affiliation{Institute of Advanced Materials (INAM), Universitat Jaume I. Av. de Vicent Sos Baynat, Castellón de la Plana, 12006 Spain}
\author{Robert Kudrawiec}
\affiliation{Department of Semiconductor Materials Engineering, Wrocław  University of Science and Technology, Wybrzeże Wyspiańskiego 27, 50-370 Wrocław, Poland}

\begin{document}

\begin{tocentry}
\includegraphics[width=8cm,height=4.45cm]{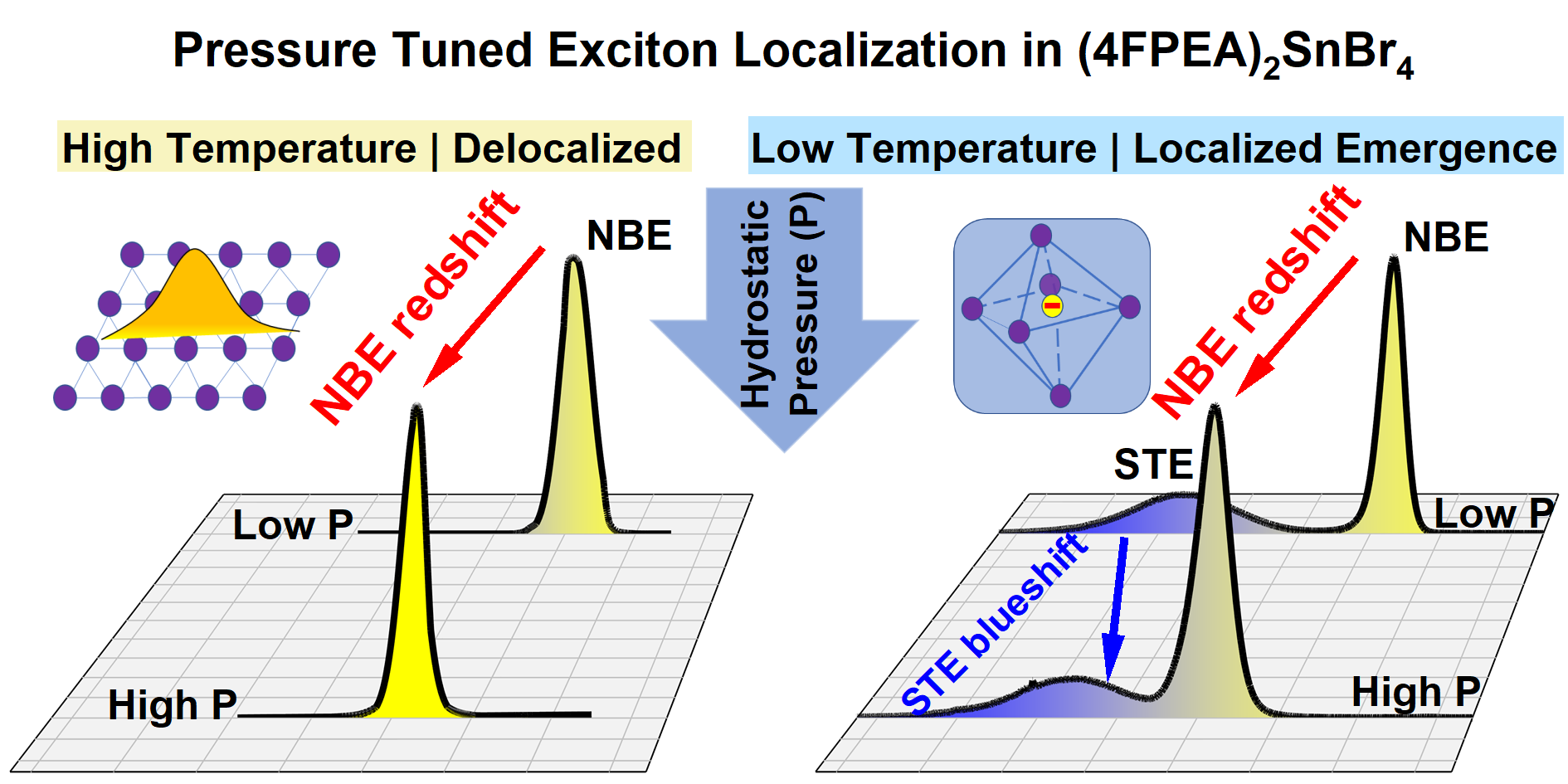}
\end{tocentry}

\begin{abstract}
Two-dimensional tin halide perovskites provide a highly tunable platform for exciton phonon coupling and local lattice distortions, enabled by their intrinsically soft lattice. We report a combined temperature and pressure dependent photoluminescence study of the layered perovskite (4FPEA)$_{2}$SnBr$_{4}$. At room temperature, its optical response is dominated by near band edge (NBE) excitons, which redshift linearly under hydrostatic pressure up to $\sim$3 GPa, indicating a rigid band edge behavior without phase transitions. Cooling reveals a broad, strongly Stokes shifted self-trapped exciton (STE) emission, evidencing a crossover from delocalized to self localized excitonic states. Strikingly, while NBE emission redshifts under pressure, STE emission exhibits an anomalous blueshift, reflecting pressure induced modification of the exciton phonon energy landscape. In contrast, the iodide analogue (4FPEA)$_{2}$SnI$_{4}$ shows no STE emission under identical conditions, highlighting the critical role of lattice rigidity and dielectric screening in stabilizing self-trapped excitons.
\end{abstract}

Tin-based halide perovskites have emerged as promising lead-free optoelectronic materials, offering reduced toxicity with strong light-matter interactions and long carrier lifetimes.\cite{1,2,3,4,5,61} Among them, two-dimensional (2D) Sn-based halide perovskites exhibit pronounced quantum confinement and enhanced excitonic localization,\cite{6,7,8} making them attractive platforms for studying exciton-phonon coupling phenomena.\cite{9,54,55,56} In these materials, the interplay between electronic structure and lattice dynamics governs key processes such as exciton self-trapping, radiative recombination, and carrier-phonon interactions.\cite{10,57,58}

In particular, the coexistence and competition between free excitons and self-trapped excitons (STE) has attracted growing attention,\cite{11,13} as STE formation is often associated with large Stokes shifts (200-500 meV), broadband emission (100-300 meV), and enhanced exciton-phonon coupling,\cite{12,14} properties that are both technologically appealing and scientifically challenging to control. Here, STEs are interpreted as small polarons, excitons that become tightly localized through strong short-range electron-phonon coupling and consequent local lattice distortion, in contrast to more delocalized large polarons associated with free excitonic state.\cite{14,60} In principle, STEs can be further categorized into those that arise intrinsically from the host lattice and those that are bound to defects or lattice imperfections. In high-quality samples where defect densities are minimal, defect-bound STE signatures are often not resolvable, leaving primarily intrinsic small polaron STE behavior in the optical response.

Exciton self-trapping in halide perovskites is generally attributed to strong coupling between electronic excitations and local lattice distortions.\cite{12} While STE emission has been widely reported in low dimensional and all inorganic halide perovskites, its emergence is highly sensitive to lattice rigidity, dielectric screening, chemical composition, and temperature.\cite{14} Despite extensive experimental and theoretical efforts, an unified understanding of how external perturbations modulate the balance between delocalized and localized excitonic state remains incomplete.\cite{15,16} Hydrostatic pressure provides a powerful, continuous, and reversible means of tuning interatomic distances, orbital overlap, and phonon spectra without introducing chemical disorder.\cite{17} Yet, pressure-dependent behavior of excitons in layered tin-based halide perovskites remain comparatively unexplored.

Notably, Sn-based halide perovskites demonstrate remarkable resilience to external stimuli such as hydrostatic pressure and temperature \cite{18,19}. These stimuli can significantly modulate crystallographic and electronic structures, leading to band gap renormalization, photoluminescence (PL) enhancement, defect-assisted recombination, or even amorphization.\cite{17} In many cases, compression activates new emissive channels or enhances STE formation,\cite{20,21,59,16} often accompanied by structural instabilities. However, such responses are highly material specific and dependent sensitively on halide chemistry, lattice stiffness and intrinsic structural softness. Understanding, the behavior of these materials under extreme perturbations is therefore important for the development of next generation perovskite-based devices,\cite{22,23} including stimuli-responsive optoelectronic, temperature sensing, mechanical stress detection and flexible electronics.\cite{24,25,26}

4-fluorophenethylammonium tin bromide ((4FPEA)$_{2}$SnBr$_{4}$) is a recently developed 2D Sn-based halide perovskite.\cite{27} While its structure allows for straightforward optoelectronic characterization, its fundamental optical properties have not yet been reported. In comparison, the closely related (PEA)$_{2}$SnBr$_{4}$ exhibits a direct band gap of $\sim$ 2.7 eV and photoluminescence around $\sim$2.64 eV,\cite{53} suggesting that similar optical behavior may be anticipated in the 4F-substituted derivative. Studies on layered Sn halide perovskites further indicate that both the halide composition and the organic spacer modulate the band gap and emission characteristics.\cite{52} Despite this, the excitonic response of (4FPEA)$_{2}$SnBr$_{4}$ under external perturbations remains largely unexplored.

\begin{figure*}[htbp!]
    \centering
    \includegraphics[width=0.9\textwidth]{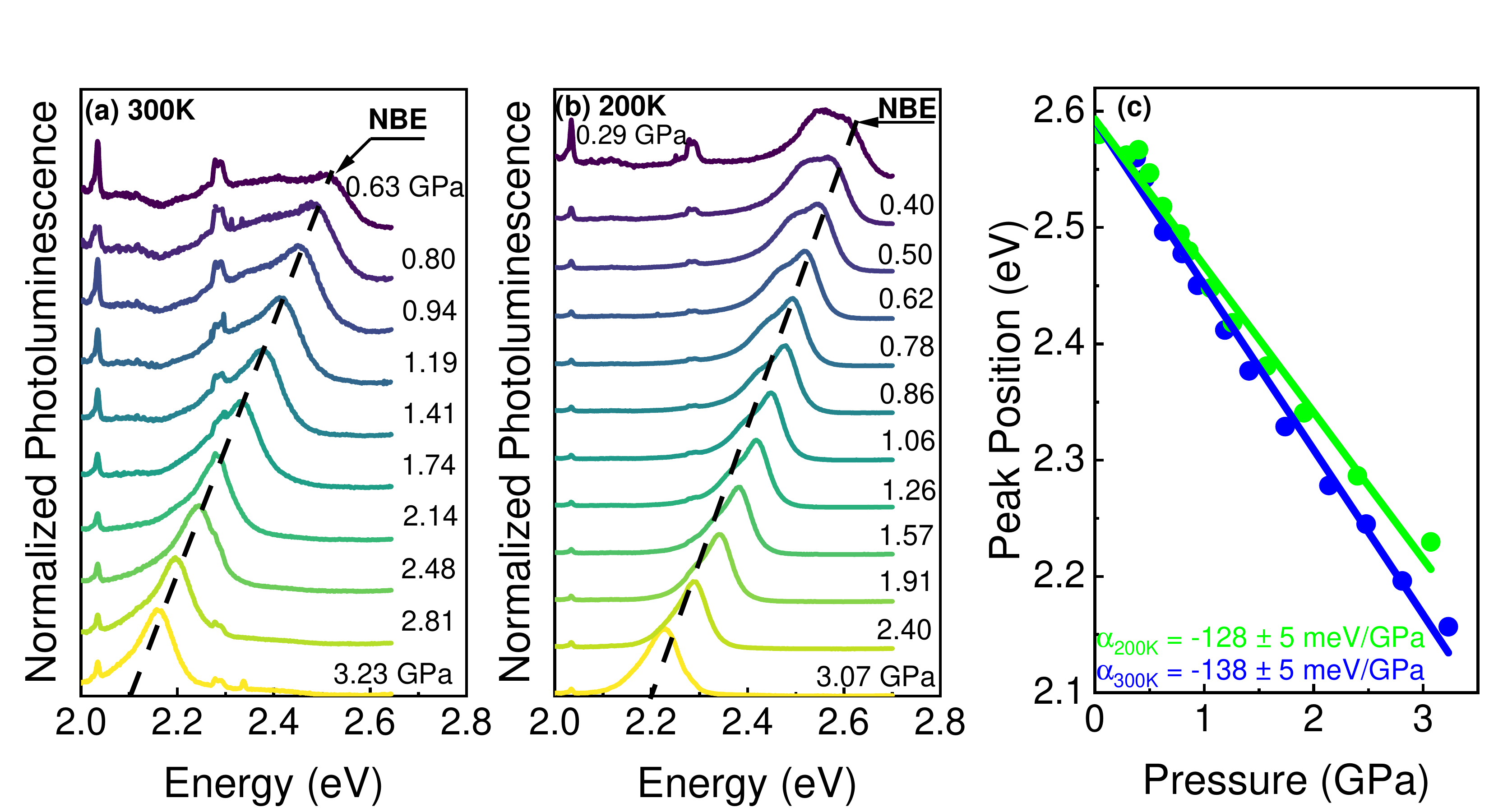}
    \caption{Normalized photoluminescence spectra of (4FPEA)$_{2}$SnBr$_{4}$ under varying hydrostatic pressure at (a) 300 and (b) 200 K. Free exciton is indicated as NBE and position changes are marked by black dashed line. (c) Pressure-induced band gap shift, extracted from the PL spectra at 300 (blue circles) and 200 K (green circles), along with corresponding linear fits (solid lines). A clear temperature dependence of the pressure coefficient is observed.}
    \label{fig:Fig1}
\end{figure*}

In this Letter, we present a comprehensive investigation of the excitonic response of (4FPEA)$_{2}$SnBr$_{4}$ under combined control of temperature and hydrostatic pressure. By tracking its PL spectra from room temperature down to cryogenic temperatures and pressures up to $\sim$ 3 GPa, we uncover a strikingly rigid high temperature excitonic response, characterized by a single near-band-edge (NBE) emission. Here, the high-energy emission close to the optical band edge is referred to as NBE excitonic emission and is commonly attributed to free exciton recombination. This emission redshifts linearly with pressure without any anomalous spectral broadening, intensity enhancement, or evidence of phase transitions what is in line with behavior reported for other perovskites. In contrast, at low temperatures a broad STE emission emerges exclusively in the bromide compound, exhibiting an anomalous blueshift under compression. Direct comparison with iodide analogues shows that STE emission is entirely suppressed in softer, more strongly screened lattices.

\begin{figure*}[htbp!]
    \centering
    \includegraphics[width=0.9\textwidth]{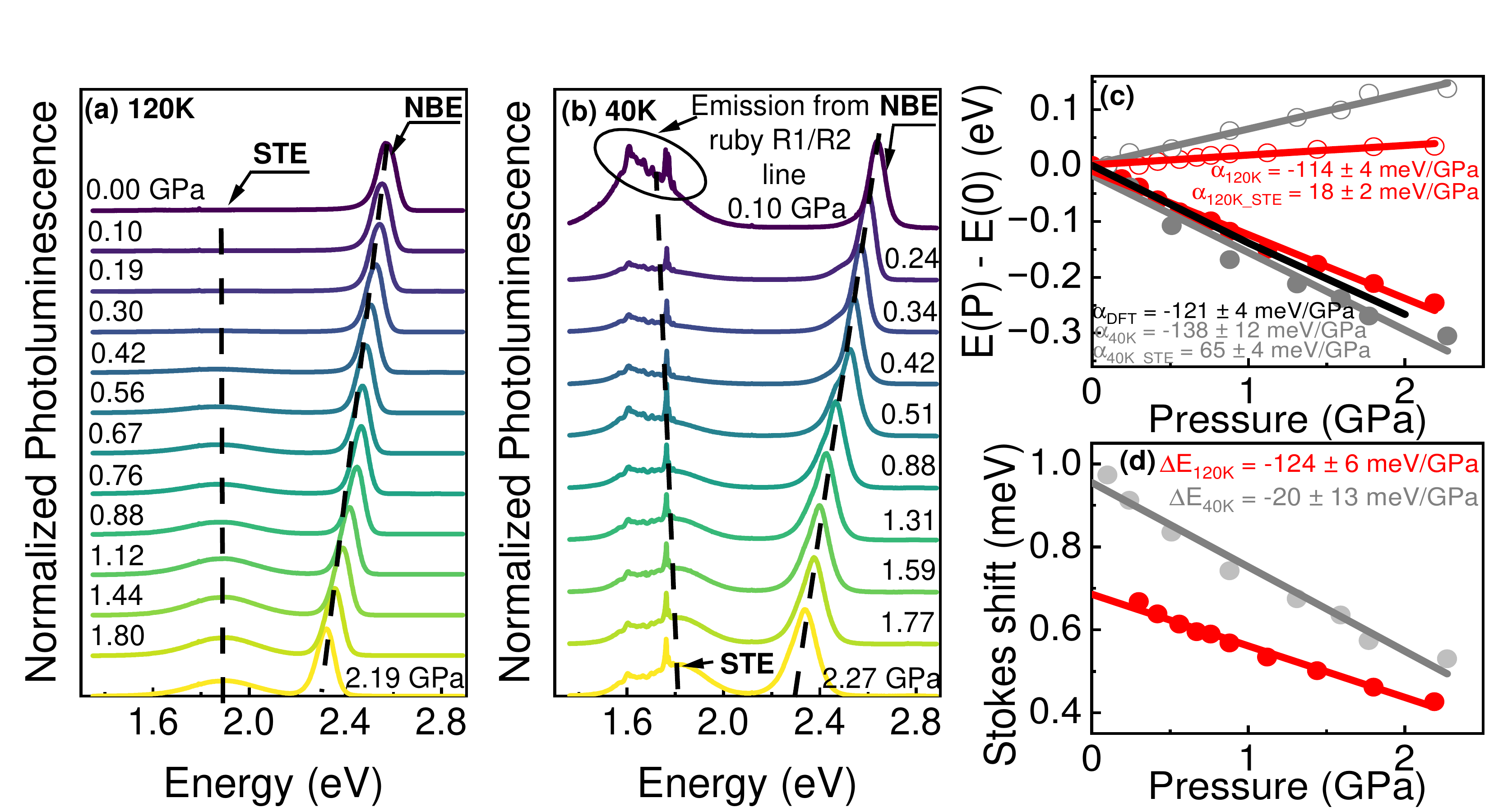}
    \caption{Pressure dependent normalized PL spectra of (4FPEA)$_{2}$SnBr$_{4}$ at (a) 120 and (b) 40 K. Self-Trapped Exciton is indicated as component STE. Positions changes are marked by black dashed lines. (c) Pressure-induced band gap shift $\Delta E = E(P) - E(P=0)$, extracted at 120 (red circles) and 40 K (grey circles). Solid circles and lines correspond to the NBE feature; open circles and solid lines represent STE feature. Black line shows DFT calculations (crystal strucutre is presented in Figure S2 in the SI). (d) Pressure dependence of the Stokes shift from the same data.}
    \label{fig:Fig2}
\end{figure*}

The synthesis of (4FPEA)$_{2}$SnBr$_{4}$ microcrystals was performed using the developed methodology elsewhere,\cite{27} as detailed in the Supporting Information (SI). To prove the formation and explore the crystallographic features of the resulting microcrystals, we performed X-ray diffraction (XRD) to identify the crystal phases obtained. Figure S1 in the SI shows the XRD pattern of the (4FPEA)$_{2}$SnBr$_{4}$ microcrystals, exhibiting high crystalinity and uniformity, showing repetitive XRD peaks with intervals of $\sim 5.4^{\circ}$, which corresponds to the interlayer distance of the inorganic layers.

Figure \ref{fig:Fig1} (a-b) presents the evolution of the normalized PL spectra of (4FPEA)$_{2}$SnBr$_{4}$ under hydrostatic pressure, measured isothermally at 300 and 200 K, respectively. Hydrostatic conditions were maintained up to approximately 3 GPa for all measurements. An assessment of the pressure-transmitting medium is provided in Figure S3 in the SI. At both temperatures, the material retains its initial structural phase throughout the investigated pressure range. The PL response is dominated by NBE emission, with no emergence of additional emissive features, consistent with ambient pressure behavior.\cite{27} Upon increasing pressure, the NBE peak exhibits a pronounced redshift, while its linewidth and relative intensity remain nearly unchanged compared to low pressures. For clarity in tracking the pressure-induced energy shifts, all spectra are normalized to their maximum intensity. A weak peaks near $\sim$2.3 and 2.0 eV arises from inadvertent illumination by a green and red light source in the experimental environment. However, this artifacts does not affect the determination of the NBE peak position. Pressure coefficients were extracted from linear fits of the NBE peak energy as a function of pressure, with peak positions obtained using Gaussian fitting (Figure \ref{fig:Fig1} (c)). The resulting coefficients are negative and slightly decrease in magnitude with increasing temperature (-138 and -128 meV/GPa at 300 and 200 K, respectively), consistent with pressure-induced band gap narrowing. This behavior reflects enhanced orbital overlap within the Sn-Br framework under compression, which lowers the band edges while preserving the excitonic character of the emission.\cite{18,32} Notably, unlike many halide perovskites, (4FPEA)$_{2}$SnBr$_{4}$ does not exhibit pressure-induced PL enhancement, band gap widening, additional emission channels, or amorphization up to $\sim$ 3 GPa.\cite{21,29,30,31} These observations indicate a comparatively rigid pressure response in which compression primarily modifies band dispersion rather than activating defect- or polaron assisted recombination pathways.\cite{33,34,59} In contrast, markedly different PL behavior is observed at lower temperatures.

\begin{figure*}[h!]
    \centering
    \includegraphics[width=0.9\textwidth]{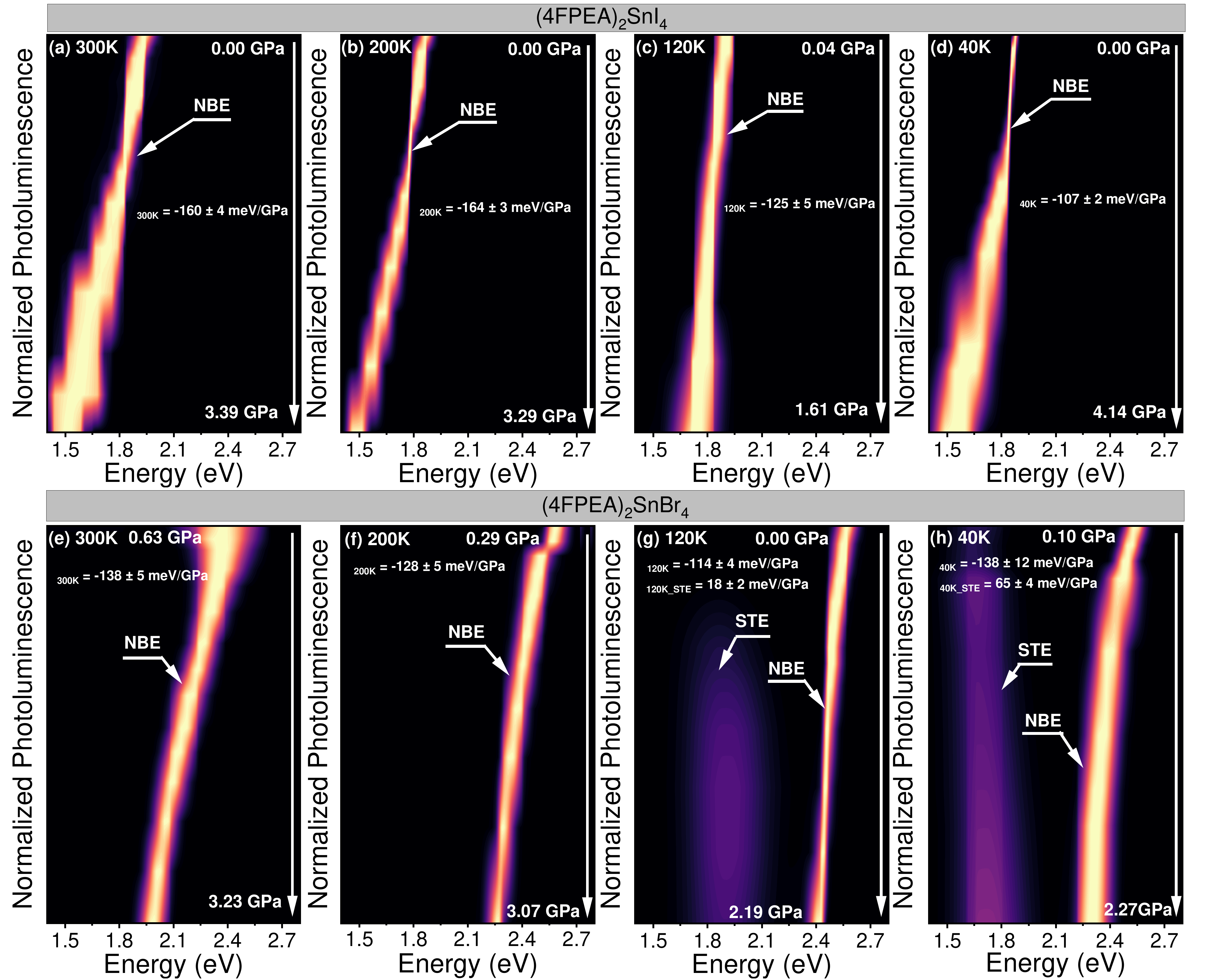}
    \caption{Pressure dependent photoluminescence spectra of two layered halide perovskites measured at variable temperature. (a-d) PL spectra of (4FPEA)$_{2}$SnI$_{4}$ collected at 300, 200, 120, and 40 K under hydrostatic pressure. (e-h) Corresponding PL spectra of (4FPEA)$_{2}$SnBr$_{4}$ acquired under indentical temperature but different pressure conditions. Bright yellow traces denote the evolution of NBE emission, whereas the violet trace indicates the spectral position of STE emission.}
    \label{fig:Fig3}
\end{figure*}

Figure \ref{fig:Fig2} (a-b) present the pressure-dependent PL spectra measured at 120 and 40 K. At both temperatures and low pressure, the spectra are dominated by NBE emission, together with a weaker and broader STE band which is separated from NBE emission by a large Stokes shift typical of STE containing small polarons.\cite{40,44} With increasing pressure, the NBE emission undergoes a pronounced redshift, whereas the STE emission exhibits a clear blueshift accompanied by a significant increase in intensity. The opposite pressure response of NBE, which can be attributed to free exciton (FX), and STE emissions highlights their distinct physical origins: while NBE emission follows the band-edge electronic structure, STE emission reflects the depth of the local lattice relaxation potential associated with exciton self-trapping.\cite{35} At 40 K, the STE emission partially overlaps with a secondary peak at 1.78 eV arising from ruby fluorescence used for pressure calibration \cite{36}. Despite this overlap, the pressure-induced blueshift of the STE emission remains unambiguous. Pressure coefficients for both NBE and STE emissions were extracted from linear fits of peak energy versus pressure, with peak positions determined by Gaussian fitting, as shown in Figure \ref{fig:Fig2} (c). Representative Gaussian fit is provided in the Figure S4 in SI. Density functional theory (DFT) results are also included for comparison. At 120 K, the NBE emission exhibits a negative pressure coefficient (-114 meV/GPa), consistent with the trend observed at higher temperatures, while the STE emission shows a positive coefficient (18 meV/GPa). At 40 K, this trend is partially disrupted by the freezing of the pressure-transmitting medium, resulting in quasi-hydrostatic conditions. In addition, the strong pressure-induced enhancement of the STE emission hampers reliable detection of the ruby R1/R2 lines, preventing precise determination of the pressure inside the diamond anvil cell. As a consequence, the NBE pressure coefficient at 40 K is reduced (-138 meV/GPa) and exhibits larger uncertainty. Nevertheless, it agrees well with the DFT predicted value (-121 meV/GPa), supporting the consistency between experiment and theory (see below Figure \ref{fig:DFT}). The STE emission at 40 K displays a what is positive pressure coefficient (65 meV/GPa), consistent with pressure-induced STE behavior reported for many halide perovskites.\cite{16,37,38,39} Up to approximately 2 GPa, no pressure-induced PL enhancement, band gap widening, or amorphization is observed. Importantly, in (4FPEA)$_{2}$SnBr$_{4}$, STE emission becomes pronounced only at low temperatures, confirming its self-trapped nature.\cite{40} Figure \ref{fig:Fig2} (d) presents the pressure dependence of the Stokes shift between NBE and STE emissions. At both temperatures, the Stokes shift decreases monotonically with increasing pressure, reflecting a progressive reduction of the lattice relaxation energy associated with exciton self-trapping. A strong temperature dependence of this effect is observed at 120 K, the Stokes shift is highly pressure sensitive, exhibiting a slope of -124 meV/GPa, whereas at 40 K the pressure induced reduction is much weaker (-20 meV/GPa). This pronounced difference indicates that thermally activated lattice degrees of freedom substantially enhance the pressure response of the NBE-STE energy separation. Upon cooling, lattice stiffening and reduced phonon activity suppress pressure driven modifications of the self-trapping potential, resulting in a markedly smaller Stokes shift variation.

\begin{figure}[htbp!]
    \centering
    \includegraphics[width=0.475\textwidth]{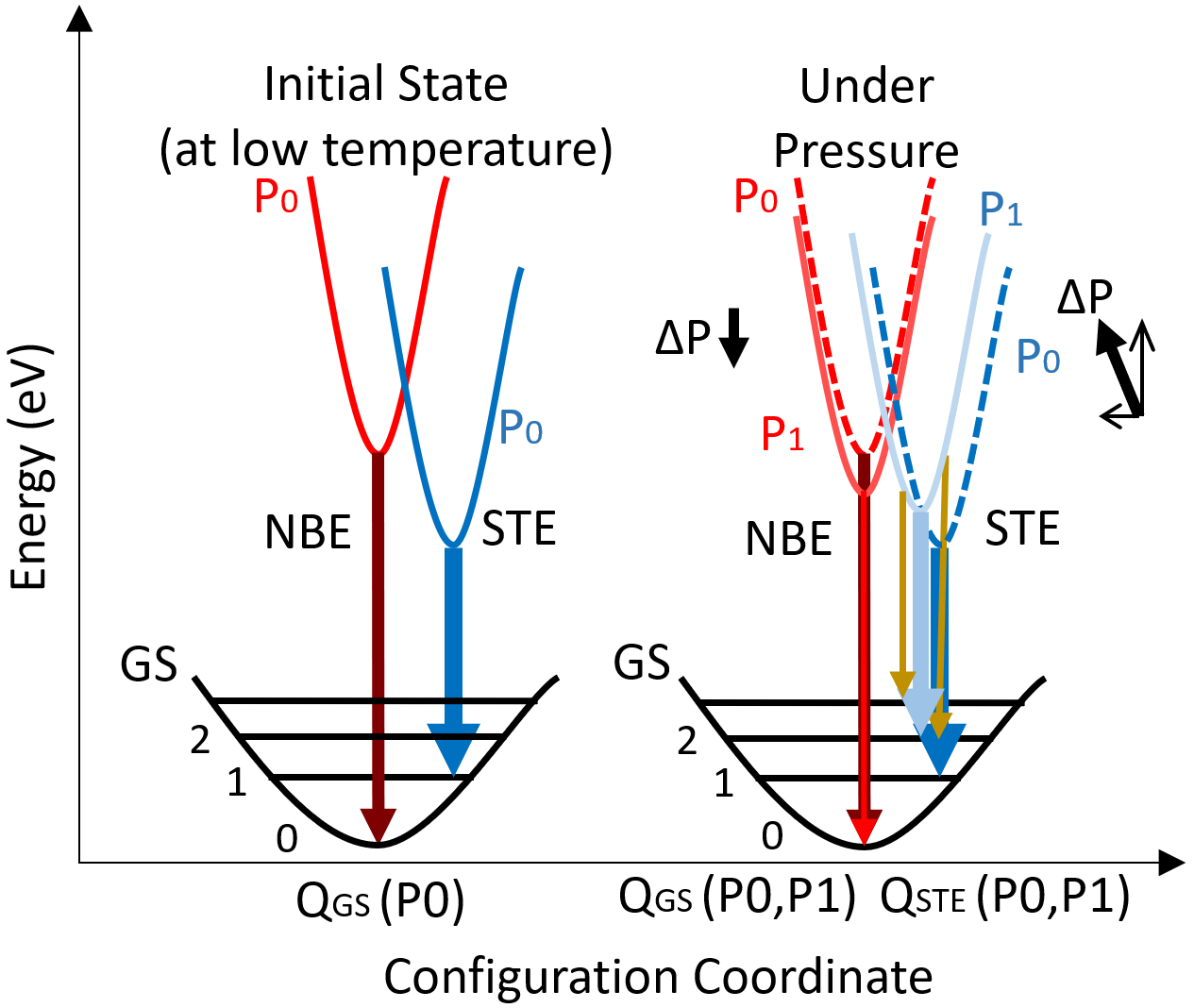}
    \caption{Schematic configuration diagram illustrating the pressure-induced redshift of the NBE emission and blueshift of STE emission in (4FPEA)$_{2}$SnBr$_{4}$. P$_{0}$ denotes ambient pressure, while P$_{1}$ larger than P$_{0}$ represents elevated pressure. GS indicates the ground state of the system.}
    \label{fig:Fig4}
\end{figure}

Temperature and pressure-dependent PL spectra of the layered perovskites (4FPEA)$_{2}$SnI$_{4}$ and (4FPEA)$_{2}$SnBr$_{4}$ shown in Figure \ref{fig:Fig3} (a-d) and (e-h), respectively, as well as it has been reported for their analogues MAPbI$_{3}$ and MAPbBr$_{3}$,\cite{18,41,42} reveal a pronounced halide-dependent exciton localization behavior. At 300 and 200 K, all compounds exhibit a single, relatively narrow emission band attributed to FX recombination, indicating that excitonic states remain predominantly delocalized at elevated temperatures.\cite{43} Upon cooling to 120 K and further to 40 K, an additional low-energy emission band emerges exclusively in the bromide-containing compounds, which we assigned in this paper to STE emission. This emission is absent in the iodide analogues over the entire investigated temperature range. The STE emission is broad ($\sim$ 350 meV) and strongly redshifted with respect to the NBE peak, displaying a large Stokes shift that points to a distinct recombination pathway.\cite{44} In addition, STE emission in the bromide perovskites shows a strong sensitivity to hydrostatic pressure. With increasing pressure, its relative intensity increases at the expense of the NBE emission, indicating efficient conversion of photoexcited carriers into localized states.\cite{45} Simultaneously, the NBE emission undergoes a gradual redshift with pressure, whereas the low-energy STE band exhibits an opposite pressure dependence and blueshifts upon compression. The contrasting pressure coefficients of these two emissions unambiguously demonstrate their different physical origins. The emergence of STE emission at low temperatures indicates that thermal fluctuations suppress stable exciton self-trapping at higher temperatures.\cite{40} The pressure-induced enhancement of STE intensity suggests stabilization of exciton localization driven by strengthened exciton-phonon coupling under lattice compression.\cite{37} At the same time, the blueshift of the STE emission under pressure can be rationalized by a reduction of the lattice relaxation energy due to pressure-induced lattice stiffening.\cite{15} The absence of STE emission in the iodide-based compounds, even at low temperatures and elevated pressures (see Figure S5 in the SI), can be rationalized by their softer lattice and stronger dielectric screening, which disfavor the formation of a stable self-trapped exciton state.\cite{46,47} While a definitive identification would require further experiments, the observed temperature and pressure-dependent PL behavior provides strong evidence consistent with STE emission in bromide perovskites, highlighting the critical role of lattice rigidity in governing exciton localization in metal-haldie perovskites.\cite{48}

Next, we explain the evolution of NBE and STE emissions with temperature and pressure using a configurational coordinate diagram,\cite{15,40,44,49,50,51} shown in Figure \ref{fig:Fig4}. In this model, NBE (red) and STE (blue) correspond to distinct energy minima along a lattice distortion coordinate, capturing the dominant exciton-phonon interactions. Optical excitation produces a delocalized NBE state, closely resembling the ground-state lattice. In bromide perovskites, strong exciton-phonon coupling drives relaxation into a shifted STE minimum, producing broad, strongly Stokes-shifted emission. Thermal activation over the barrier separating NBE and STE states governs their relative populations, explaining why STE emission dominates at low temperatures. Hydrostatic pressure reshapes the energy landscape, promoting exciton self-trapping and stiffening the lattice, which enhances the STE population. Notably, we observe that the Stokes shift decreases with increasing pressure. This behavior can be attributed to a reduction in the lattice relaxation energy, reflecting a pressure induced change in the energy formation of small polarons. Consequently, STE emission exhibits a pressure induced blueshift, in contrast to the gradual redshift of NBE emission. In iodide analogues, STE emission is absent (see Figure S5 in SI showing PL spectrum of (4FPEA)$_{2}$SnI$_{4}$ at 40 K and ambient pressure (0 GPa) plotted on a logarithmic scale), as the softer lattice and stronger dielectric screening suppress exciton self-trapping. Displacement along the configurational coordinate reflects lattice relaxation energy, which can be quantitatively inferred from the Stokes shift. These observations underscore the critical role of lattice rigidity and exciton-phonon coupling in controlling exciton self-trapping, while also revealing how external pressure can be used to tune polaron formation and the associated emission properties in layered halide perovskites.

\begin{figure}[htbp!]
    \centering
    \includegraphics[width=0.45\textwidth]{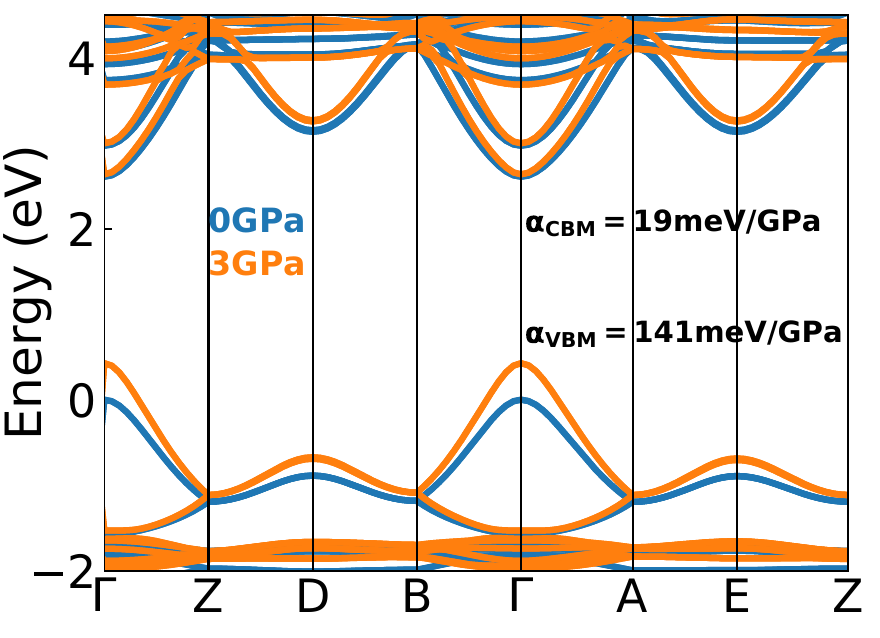}
    \caption{Pressure-dependent band structures of (4FPEA)$_{2}$SnBr$_{4}$ at 0 GPa and 3 GPa, revealing a persistent direct band gap and a significant pressure induced redshift. Values of pressure coefficients for VBM and CBM are also given.}
    \label{fig:DFT}
\end{figure}

To contextualize the experimental pressure response within an electronic structure framework, we performed DFT calculations of the pressure-dependent band structure of (4FPEA)$_{2}$SnBr$_{4}$. As shown in Figure \ref{fig:DFT}, the material retains a direct band gap up to 3 GPa, demonstrating the robustness of the band edge under moderate hydrostatic compression. Increasing pressure induces a pronounced band gap reduction, consistent with enhanced orbital overlap and lattice compression within the inorganic Sn-Br framework. The calculated band gap pressure coefficient (-121 meV/GPa) closely matches the experimentally extracted NBE pressure coefficient, providing strong evidence that the pressure evolution of the NBE emission is governed predominantly by band edge renormalization. By contrast, the exciton binding energy exhibits only a weak pressure dependence, indicating that changes in excitonic Coulomb interactions play a secondary role. This clear separation of energy scales demonstrates that the anomalous pressure response of the STE emission cannot originate from band edge effects, but instead arises from pressure-induced modifications of lattice relaxation and exciton self-trapping energetics, as captured by the configurational-coordinate model.

In conclusion, we have demonstrated (4FPEA)$_{2}$SnBr$_{4}$ as a model system for elucidating the interplay between lattice dynamics and exciton localization in 2D tin halide perovskites. The contrasting pressure responses of NBE and STE reveal that exciton phonon coupling and lattice rigidity can be systematically tuned to control polaron formation and energy relaxation pathways. The absence of STE in iodide analogue (4FPEA)$_{2}$SnI$_{4}$ underscores the delicate balance between lattice stiffness and dielectric screening in stabilizing self-trapped states. More broadly, our findings demonstrate that external perturbations such as hydrostatic pressure can serve as a precise probe of polaronic phenomena, advancing fundamental understanding and guiding the rational design of next generation lead free perovskite materials.

\begin{acknowledgement}
This work was financed by the National Science Centre (NC) Poland under OPUS 29 grant no. 2025/57/B/ST3/03683. This work is partially funded by the Ministerio de Ciencia e Innovación of the Spanish Government by the project PLEDs, PID2022-140090OB-C21/AEI/10.13039/501100011033/FEDER. Calculations have been carried out using resources provided by Wroclaw Centre for Networking and Supercomputing (https://wcss.pl)
\end{acknowledgement}

\begin{suppinfo}
Detailed materials and methods, experimental layout, 4FPEA molecular structure. Comment about hydrostaticity of Daphne 7575. Example of Gaussian fitting. Photoluminescence spectrum of (4FPEA)$_{2}$SnI$_{4}$ and (4FPEA)$_{2}$SnBr$_{4}$ at 40 K and
ambient pressure plotted on a logarithmic intensity scale.
\end{suppinfo}

%\nocite{*}
%\section*{References}
%\bibliographystyle{prsty}
%\bibliographystyle{apsrev4-2.bst}
\bibliography{Ref}

@article{1,
author = {Jokar, Efat and Cai, Liang and Han, Jiye and Nacpil, Edric John Cruz and Jeon, Il},
title = {Emerging Opportunities in Lead-Free and Lead–Tin Perovskites for Environmentally Viable Photodetector Applications},
journal = {Chemistry of Materials},
volume = {35},
number = {9},
pages = {3404-3426},
year = {2023},
doi = {10.1021/acs.chemmater.3c00345},
}

@Article{2,
author ="Li, Xiao-Zhen and Ye, Yilong and Cao, Yu and Zhang, Diwei and Lin, Yuan and Chang, Jin and Zhu, Lin and Wang, Nana and Huang, Wei and Wang, Jianpu",
title  ="Tin-halide perovskites for light-emitting diodes",
journal  ="Chem. Soc. Rev.",
year  ="2025",
volume  ="54",
issue  ="14",
pages  ="6697-6725",
publisher  ="The Royal Society of Chemistry",
doi  ="10.1039/D5CS00340G",
}

@article{3,
author = {Abate, Antonio},
title = {Stable Tin-Based Perovskite Solar Cells},
journal = {ACS Energy Letters},
volume = {8},
number = {4},
pages = {1896-1899},
year = {2023},
doi = {10.1021/acsenergylett.3c00282},
}

@article{4,
  author  = {Zhao, Dewei and Yu, Yue and Wang, Changlei and Liao, Weiqiang and
             Shrestha, Niraj and Grice, Corey R. and Cimaroli, Alexander J. and
             Guan, Lei and Ellingson, Randy J. and Zhu, Kai and
             Zhao, Xingzhong and Xiong, Ren-Gen and Yan, Yanfa},
  title   = {Low-bandgap mixed tin--lead iodide perovskite absorbers with long carrier lifetimes for all-perovskite tandem solar cells},
  journal = {Nature Energy},
  year    = {2017},
  volume  = {2},
  number  = {4},
  pages   = {17018},
  doi     = {10.1038/nenergy.2017.18},
  issn    = {2058-7546},
}

@article{5,
author = {Li, Yusheng and Wang, Dandan and Yang, Yongge and Ding, Chao and Hu, Yuyu and Liu, Feng and Wei, Yuyao and Liu, Dong and Li, Hua and Shi, Guozheng and Chen, Shikai and Li, Hongshi and Fuchimoto, Akihito and Tosa, Keita and Hiroki, Unno and Hayase, Shuzi and Wei, Huiyun and Shen, Qing},
title = {Stable Inorganic Colloidal Tin and Tin–Lead Perovskite Nanocrystals with Ultralong Carrier Lifetime via Sn(IV) Control},
journal = {Journal of the American Chemical Society},
volume = {146},
number = {5},
pages = {3094-3101},
year = {2024},
doi = {10.1021/jacs.3c10060},
}

@article{6,
  title = {Quantum Confinement and Dielectric Deconfinement in Quasi-Two-Dimensional Perovskites: Their Roles in Light-Emitting Diodes},
  author = {Chakraborty, Raja and Paul, Goutam and Pal, Amlan J.},
  journal = {Phys. Rev. Appl.},
  volume = {17},
  issue = {5},
  pages = {054045},
  numpages = {10},
  year = {2022},
  month = {May},
  publisher = {American Physical Society},
  doi = {10.1103/PhysRevApplied.17.054045},
}

@article{7,
author = {Hansen, Kameron R. and McClure, C. Emma and Powell, Daniel and Hsieh, Hao-Chieh and Flannery, Laura and Garden, Kelsey and Miller, Edwin J. and King, Daniel J. and Sainio, Sami and Nordlund, Dennis and Colton, John S. and Whittaker-Brooks, Luisa},
title = {Low Exciton Binding Energies and Localized Exciton–Polaron States in 2D Tin Halide Perovskites},
journal = {Advanced Optical Materials},
volume = {10},
number = {9},
pages = {2102698},
doi = {https://doi.org/10.1002/adom.202102698},
year = {2022}
}

@article{8,
author = {Chen, Yameng and Wang, Zhaoyu and Wei, Youchao and Liu, Yongsheng and Hong, Maochun},
title = {Exciton Localization for Highly Luminescent Two-Dimensional Tin-Based Hybrid Perovskites through Tin Vacancy Tuning},
journal = {Angewandte Chemie International Edition},
volume = {62},
number = {18},
pages = {e202301684},
doi = {https://doi.org/10.1002/anie.202301684},
year = {2023}
}

@article{9,
author  = {Zhang, Tianju and Zhou, Chaocheng and Feng, Xuezhen and
             Dong, Ningning and Chen, Hong and Chen, Xianfeng and
             Zhang, Long and Lin, Jia and Wang, Jun},
title   = {Regulation of the luminescence mechanism of two-dimensional tin halide perovskites},
journal = {Nature Communications},
year    = {2022},
volume  = {13},
number  = {1},
pages   = {60},
doi     = {10.1038/s41467-021-27663-0},
}

@article{10,
author = {Duan, Jianing and Li, Jingrui and Divitini, Giorgio and Cortecchia, Daniele and Yuan, Fang and You, Jiaxue and Liu, Shengzhong (Frank) and Petrozza, Annamaria and Wu, Zhaoxin and Xi, Jun},
title = {2D Hybrid Perovskites: From Static and Dynamic Structures to Potential Applications},
journal = {Advanced Materials},
volume = {36},
number = {30},
pages = {2403455},
doi = {https://doi.org/10.1002/adma.202403455},
year = {2024}
}

@article{11,
author = {Gualdrón-Reyes, Andrés F.},
title = {Self–Trapped Exciton versus Band–Edge Electron Transitions: Insights of the Factors Affecting the Optical Properties of Lead–Free Sn–Halide Perovskites},
journal = {Advanced Optical Materials},
volume = {13},
number = {2},
pages = {2402043},
doi = {https://doi.org/10.1002/adom.202402043},
year = {2025}
}

@article{12,
author  = {Li, Junze and Wang, Haizhen and Li, Dehui},
title   = {Self-trapped excitons in two-dimensional perovskites},
journal = {Frontiers of Optoelectronics},
year    = {2020},
volume  = {13},
number  = {3},
pages   = {225--234},
doi     = {10.1007/s12200-020-1051-x},
}

@article{13,
author = {Chen, Yameng and Wang, Zhaoyu and Wei, Youchao and Liu, Yongsheng and Hong, Maochun},
title = {Exciton Localization for Highly Luminescent Two-Dimensional Tin-Based Hybrid Perovskites through Tin Vacancy Tuning},
journal = {Angewandte Chemie International Edition},
volume = {62},
number = {18},
pages = {e202301684},
doi = {https://doi.org/10.1002/anie.202301684},
year = {2023}
}

@article{14,
author = {Li, Shunran and Luo, Jiajun and Liu, Jing and Tang, Jiang},
title = {Self-Trapped Excitons in All-Inorganic Halide Perovskites: Fundamentals, Status, and Potential Applications},
journal = {The Journal of Physical Chemistry Letters},
volume = {10},
number = {8},
pages = {1999-2007},
year = {2019},
doi = {10.1021/acs.jpclett.8b03604},
}

@article{15,
title = {Theory and experiments of pressure-tunable broadband light emission from self-trapped excitons in metal halide crystals},
journal = {Materials Today Physics},
volume = {30},
pages = {100926},
year = {2023},
issn = {2542-5293},
doi = {https://doi.org/10.1016/j.mtphys.2022.100926},
author = {Shenyu Dai and Xinxin Xing and Viktor G. Hadjiev and Zhaojun Qin and Tian Tong and Guang Yang and Chong Wang and Lijuan Hou and Liangzi Deng and Zhiming Wang and Guoying Feng and Jiming Bao},
}

@article{16,
author = {Mączka, Mirosław and Sobczak, Szymon and Roszak, Kinga and Vasconcelos, Daniel Linhares Militão and Dybała, Filip and Herman, Artur P. and Kudrawiec, Robert and Katrusiak, Andrzej and Freire, Paulo T. C.},
title = {Pressure-Induced Detrapping from Self-Trapped Excitons to Free Excitons toward Enhanced Emission and Piezochromism in Ruddlesden–Popper (110)-Oriented Perovskites},
journal = {ACS Applied Materials \& Interfaces},
volume = {17},
number = {42},
pages = {58452-58466},
year = {2025},
doi = {10.1021/acsami.5c14096},
}

@article{17,
author = {Jaffe, Adam and Lin, Yu and Karunadasa, Hemamala I.},
title = {Halide Perovskites under Pressure: Accessing New Properties through Lattice Compression},
journal = {ACS Energy Letters},
volume = {2},
number = {7},
pages = {1549-1555},
year = {2017},
doi = {10.1021/acsenergylett.7b00284},
}

@article{18,
author = {Bartoszewicz, Rafał and Ziembicki, Jakub and Zdanowicz, Ewelina and Herman, Artur P. and Serafińczuk, Jarosław and Sánchez-Diaz, Jesús and Das Adhikari, Samrat and Mora-Seró, Iván and Kudrawiec, Robert},
title = {Giant Band Gap Narrowing under Hydrostatic Pressure in (4FP)2SnI4 Halide Perovskite},
journal = {The Journal of Physical Chemistry Letters},
volume = {16},
number = {25},
pages = {6372-6377},
year = {2025},
doi = {10.1021/acs.jpclett.5c00903},
}

@article{19,
author = {Kong, Lingping and Gong, Jue and Spanopoulos, Ioannis and Yan, Shuai and Li, Zhongyang and Zhu, Zhikai and Liu, Xingyi and Zhu, Yinning and Dong, Hongliang and Shu, Haiyun and Hu, Qingyang and Yang, Wenge and Mao, Ho-kwang and Kanatzidis, Mercouri G. and Liu, Gang},
title = {Revealing the Universal Pressure-Driven Behavior of Hybrid Halide Perovskites and Unique Optical Modifiability in Extremely Soft 2D Tin-Based System},
journal = {Advanced Functional Materials},
volume = {34},
number = {46},
pages = {2414437},
doi = {https://doi.org/10.1002/adfm.202414437},
year = {2024}
}

@article{20,
author = {Fang, Ya-Si and Guo, Yi-Fan and Xu, Hao-Jin and Chen, Xiao-Gang and Zhou, Lin and Wang, Zhong-Xia and Tang, Yuan-Yuan and Qin, Yan},
title = {Pressure-Modulated Near-Infrared Photoluminescence of a Two-Dimensional Germanium Halide Perovskite},
journal = {The Journal of Physical Chemistry Letters},
volume = {16},
number = {39},
pages = {10126-10133},
year = {2025},
doi = {10.1021/acs.jpclett.5c02131},
}

@article{21,
author = {Zhao, Wenya and Xiao, Guanjun and Zou, Bo},
title = {Pressure-induced emission (PIE) in halide perovskites toward promising applications in scintillators and solid-state lighting},
journal = {Aggregate},
volume = {5},
number = {1},
pages = {e461},
doi = {https://doi.org/10.1002/agt2.461},
year = {2024}
}

@article{22,
author    = {Zhumekenov, Ayan A. and Saidaminov, Makhsud I. and Mohammed, Omar F. and Bakr, Osman M.},
title     = {Stimuli-responsive switchable halide perovskites: Taking advantage of instability},
journal   = {Joule},
year      = {2021},
volume    = {5},
number    = {8},
pages     = {2027--2046},
doi       = {10.1016/j.joule.2021.07.008},
publisher = {Elsevier},
}

@article{23,
author    = {Sun, Yuqi and Ge, Lishuang and Dai, Linjie and Cho, Changsoon and Ferrer Orri, Jordi and Ji, Kangyu and Zelewski, Szymon J. and Liu, Yun and Mirabelli, Alessandro J. and Zhang, Youcheng and Huang, Jun-Yu and Wang, Yusong and Gong, Ke and Lai, May Ching and Zhang, Lu and Yang, Dan and Lin, Jiudong and Tennyson, Elizabeth M. and Ducati, Caterina and Stranks, Samuel D. and Cui, Lin-Song and Greenham, Neil C.},
title     = {Bright and stable perovskite light-emitting diodes in the near-infrared range},
journal   = {Nature},
year      = {2023},
volume    = {615},
number    = {7954},
pages     = {830--835},
doi       = {10.1038/s41586-023-05792-4},
}

@article{24,
	title = {Pressure engineering of photovoltaic perovskites},
	volume = {27},
	issn = {1369-7021},
	url = {https://www.sciencedirect.com/science/article/pii/S1369702118309489},
	doi = {10.1016/j.mattod.2019.02.016},
	urldate = {2025-02-15},
	journal = {Materials Today},
	author = {Liu, Gang and Kong, Lingping and Yang, Wenge and Mao, Ho-kwang},
	month = jul,
	year = {2019},
	pages = {91--106},
}

@article{25,
	title = {Development of {Perovskite} {Quantum} {Dots} for {Two}-{Dimensional} {Temperature} {Sensors}},
	volume = {6},
	url = {https://doi.org/10.1021/acsanm.3c00144},
	doi = {10.1021/acsanm.3c00144},
	number = {6},
	urldate = {2025-02-15},
	journal = {ACS Applied Nano Materials},
	author = {Zhu, Yanshen and Buitenhuis, Johan and Förster, Beate and Vetrano, Maria Rosaria and Koos, Erin},
	month = mar,
	year = {2023},
	pages = {4661--4671},
}

@article{26,
author = {Vescio, Giovanni and Dirin, Dmitry N. and González-Torres, Sergio and Sanchez-Diaz, Jesús and Vidal, Rosario and Franco, Iván P. and Adhikari, Samrat Das and Chirvony, Vladimir S. and Martínez-Pastor, Juan P. and Vinocour Pacheco, Felipe A. and Przypis, Lukasz and Öz, Senol and Hernández, Sergi and Cirera, Albert and Mora-Seró, Iván and Kovalenko, Maksym V. and Garrido, Blas},
title = {Inkjet-Printed Red-Emitting Flexible LEDs Based on Sustainable Inks of Layered Tin Iodide Perovskite},
journal = {Advanced Sustainable Systems},
volume = {8},
number = {9},
pages = {2400060},
doi = {https://doi.org/10.1002/adsu.202400060},
year = {2024}
}

@article{27,
author = {Sanchez-Diaz, Jesus and Rodriguez-Pereira, Jhonatan and Das Adhikari, Samrat and Mora-Seró, Iván},
title = {Synthesis of Hybrid Tin-Based Perovskite Microcrystals for LED Applications},
journal = {Advanced Science},
volume = {11},
number = {34},
pages = {2403835},
doi = {https://doi.org/10.1002/advs.202403835},
year = {2024}
}

@Article{29,
author ="Wu, Kewei and Bera, Ashok and Ma, Chun and Du, Yuanmin and Yang, Yang and Li, Liang and Wu, Tom",
title  ="Temperature-dependent excitonic photoluminescence of hybrid organometal halide perovskite films",
journal  ="Phys. Chem. Chem. Phys.",
year  ="2014",
volume  ="16",
issue  ="41",
pages  ="22476-22481",
publisher  ="The Royal Society of Chemistry",
doi  ="10.1039/C4CP03573A",
}

@article{30,
  title = {Pressure-induced emission enhancement and bandgap narrowing: Experimental investigations and first-principles theoretical simulations on the model halide perovskite ${\mathrm{Cs}}_{3}{\mathrm{Sb}}_{2}{\mathrm{Br}}_{9}$},
  author = {Samanta, Debabrata and Chaudhary, Sonu Pratap and Ghosh, Bishnupada and Bhattacharyya, Sayan and Shukla, Gaurav and Mukherjee, Goutam Dev},
  journal = {Phys. Rev. B},
  volume = {105},
  issue = {10},
  pages = {104103},
  numpages = {11},
  year = {2022},
  month = {Mar},
  publisher = {American Physical Society},
  doi = {10.1103/PhysRevB.105.104103},
}

@article{31,
author = {Lü, Xujie and Wang, Yonggang and Stoumpos, Constantinos C. and Hu, Qingyang and Guo, Xiaofeng and Chen, Haijie and Yang, Liuxiang and Smith, Jesse S. and Yang, Wenge and Zhao, Yusheng and Xu, Hongwu and Kanatzidis, Mercouri G. and Jia, Quanxi},
title = {Enhanced Structural Stability and Photo Responsiveness of CH3NH3SnI3 Perovskite via Pressure-Induced Amorphization and Recrystallization},
journal = {Advanced Materials},
volume = {28},
number = {39},
pages = {8663-8668},
doi = {https://doi.org/10.1002/adma.201600771},
year = {2016}
}

@article{32,
title = {Origin of pressure-induced band gap tuning in tin halide perovskites††Electronic supplementary information (ESI) available: Experimental and computational details. See DOI: 10.1039/d0ma00731e},
journal = {Materials Advances},
volume = {1},
number = {8},
pages = {2840-2845},
year = {2020},
issn = {2633-5409},
doi = {https://doi.org/10.1039/d0ma00731e},
author = {Mauro Coduri and Thomas B. Shiell and Timothy A. Strobel and Arup Mahata and Federico Cova and Edoardo Mosconi and Filippo {De Angelis} and Lorenzo Malavasi},
}

@Article{33,
author ="Yang, Bowen and Bogachuk, Dmitry and Suo, Jiajia and Wagner, Lukas and Kim, Hobeom and Lim, Jaekeun and Hinsch, Andreas and Boschloo, Gerrit and Nazeeruddin, Mohammad Khaja and Hagfeldt, Anders",
title  ="Strain effects on halide perovskite solar cells",
journal  ="Chem. Soc. Rev.",
year  ="2022",
volume  ="51",
issue  ="17",
pages  ="7509-7530",
publisher  ="The Royal Society of Chemistry",
doi  ="10.1039/D2CS00278G",
}

@article{34,
author = {Wu, Jinpeng and Liu, Shun-Chang and Li, Zongbao and Wang, Shuo and Xue, Ding-Jiang and Lin, Yuan and Hu, Jin-Song},
title = {Strain in perovskite solar cells: origins, impacts and regulation},
journal = {National Science Review},
volume = {8},
number = {8},
pages = {nwab047},
year = {2021},
month = {03},
issn = {2095-5138},
doi = {10.1093/nsr/nwab047},
}

@article{35,
author    = {Bin Xu and Yawen Li and Peibin Hong and Peijie Zhang and Jiang Han and Zewen Xiao and Zewei Quan},
title     = {Pressure-controlled free exciton and self-trapped exciton emission in quasi-one-dimensional hybrid lead bromides},
journal   = {Nature Communications},
year      = {2024},
volume    = {15},
number    = {1},
pages     = {7403},
doi       = {10.1038/s41467-024-51836-2},
}

@article{36,
author = {F. Datchi and A. Dewaele and P. Loubeyre and R. Letoullec and Y. Le Godec and B. Canny},
title = {Optical pressure sensors for high-pressure–high-temperature studies in a diamond anvil cell},
journal = {High Pressure Research},
volume = {27},
number = {4},
pages = {447--463},
year = {2007},
publisher = {Taylor \& Francis},
doi = {10.1080/08957950701659593},
}

@Article{37,
author ="Jing, Xiaoling and Sun, Rui and Tian, Hui and Liu, Ran and Liu, Bo and Zhou, Donglei and Li, Quanjun and Liu, Bingbing",
title  ="Evolution of self-trapped exciton emission tuned by high pressure in 2D all-inorganic cesium lead halide nanosheets",
journal  ="J. Mater. Chem. C",
year  ="2022",
volume  ="10",
issue  ="22",
pages  ="8711-8718",
publisher  ="The Royal Society of Chemistry",
doi  ="10.1039/D2TC01465C",
}

@article{38,
author = {Gao, Fei-Fei and Qin, Yan and Li, Zhi-Gang and Li, Wei and Hao, Jing and Li, Xiang and Liu, Yungui and Howard, Christopher J. and Wu, Xiang and Jiang, Xingxing and Lin, Zheshuai and Lu, Peixiang and Bu, Xian-He},
title = {Unusual Pressure-Induced Self-Trapped Exciton to Free Exciton Transfer in Chiral 2D Lead Bromide Perovskites},
journal = {ACS Nano},
volume = {18},
number = {4},
pages = {3251-3259},
year = {2024},
doi = {10.1021/acsnano.3c09756},
}

@article{39,
author = {Shi, Han and Chen, Lin and Moutaabbid, Hicham and Feng, Zhenbao and Zhang, Guozhao and Wang, Lingrui and Li, Yinwei and Guo, Haizhong and Liu, Cailong},
title = {Mechanism of Pressure-Modulated Self-Trapped Exciton Emission in Cs2TeCl6 Double Perovskite},
journal = {Small},
volume = {20},
number = {48},
pages = {2405692},
doi = {https://doi.org/10.1002/smll.202405692},
year = {2024}
}

@Article{40,
author ="Han, Ying and Cheng, Xiaohua and Cui, Bin-Bin",
title  ="Factors influencing self-trapped exciton emission of low-dimensional metal halides",
journal  ="Mater. Adv.",
year  ="2023",
volume  ="4",
issue  ="2",
pages  ="355-373",
publisher  ="RSC",
doi  ="10.1039/D2MA00676F",
}

@article{41,
author = {Pieniążek, Agnieszka and Dybała, Filip and Polak, Maciej P. and Przypis, $\L{}$ukasz and Herman, Artur P. and Kopaczek, Jan and Kudrawiec, Robert},
title = {Bandgap Pressure Coefficient of a CH3NH3PbI3 Thin Film Perovskite},
journal = {The Journal of Physical Chemistry Letters},
volume = {14},
number = {28},
pages = {6470-6476},
year = {2023},
doi = {10.1021/acs.jpclett.3c01258},
}

@article{42,
author = {Pieniążek, Agnieszka and Dybała, Filip and Przypis, $\L{}$ukasz and Polak, Maciej P. and Norek, Małgorzata and Kowalski, Bogdan J. and Kudrawiec, Robert},
title = {Beyond the Cubic Phase: Pressure-Induced Bandgap Modulation in a CH3NH3PbBr3 Perovskite at Low Temperatures},
journal = {Advanced Optical Materials},
volume = {14},
number = {5},
pages = {e03177},
doi = {https://doi.org/10.1002/adom.202503177},
year = {2026}
}

@article{43,
doi = {10.1088/1674-1056/ad028e},
url = {https://doi.org/10.1088/1674-1056/ad028e},
year = {2023},
month = {nov},
publisher = {Chinese Physical Society and IOP Publishing Ltd},
volume = {32},
number = {11},
pages = {117802},
author = {Tan, Jiayu and Zhou, Yixuan and Lu, De and Feng, Xukun and Liu, Yuqi and Zhang, Mengen and Lu, Fangzhengyi and Huang, Yuanyuan and Xu, Xinlong},
title = {Temperature-dependent photoluminescence of lead-free cesium tin halide perovskite microplates},
journal = {Chinese Physics B},
}

@article{44,
author = {Li, Shunran and Luo, Jiajun and Liu, Jing and Tang, Jiang},
title = {Self-Trapped Excitons in All-Inorganic Halide Perovskites: Fundamentals, Status, and Potential Applications},
journal = {The Journal of Physical Chemistry Letters},
volume = {10},
number = {8},
pages = {1999-2007},
year = {2019},
doi = {10.1021/acs.jpclett.8b03604},
}

@article{45,
author = {Rong, Yan and Wang, Lingrui and Wang, Yongheng and Wang, Wenxin and Wang, Jiaxiang and Yuan, Yifang and Shahzadi, Urooj and Chen, Junnian and Zhang, Lei and Wang, Kai and Guo, Haizhong},
title = {Pressure-Induced Unusual Transition of Luminescence Mechanics from Self-Trapped Exciton to Free Exciton Emission in Lead Bromide Perovskitoids},
journal = {Advanced Optical Materials},
volume = {13},
number = {3},
pages = {2402086},
doi = {https://doi.org/10.1002/adom.202402086},
year = {2025}
}

@article{46,
author = {Gautier, Romain and Paris, Michael and Massuyeau, Florian},
title = {Exciton Self-Trapping in Hybrid Lead Halides: Role of Halogen},
journal = {Journal of the American Chemical Society},
volume = {141},
number = {32},
pages = {12619-12623},
year = {2019},
doi = {10.1021/jacs.9b04262},
}

@Article{47,
AUTHOR = {Xing, Yifeng and Yin, Jialin and Qiao, Yifei and Zhao, Jie and He, Haiyang and Zhao, Danyang and Zhang, Wanlu and Mei, Shiliang and Guo, Ruiqian},
TITLE = {Gram-Scale Synthesis and Optical Properties of Self-Trapped-Exciton-Emitting Two-Dimensional Tin Halide Perovskites},
JOURNAL = {Nanomaterials},
VOLUME = {15},
YEAR = {2025},
NUMBER = {11},
ARTICLE-NUMBER = {818},
DOI = {10.3390/nano15110818}
}

@article{48,
author = {Liang, Yin and Jiang, Yingjie and Du, Ke-Zhao and Lin, Yang-Peng and Ma, Xinyuan and Qiu, Daping and Wang, Ziyu and Hou, Yanglong and Wei, Xiaoding and Zhang, Qing},
title = {A High-Rigidity Organic–Inorganic Metal Halide Hybrid Enabling Reversible and Enhanced Self-Trapped Exciton Emission under High Pressure},
journal = {Nano Letters},
volume = {23},
number = {16},
pages = {7599-7606},
year = {2023},
doi = {10.1021/acs.nanolett.3c02205},
}

@article{49,
author = {Zhao, Wenya and Fu, Ruijing and Yang, Jiayi and Xiao, Guanjun and Zou, Bo},
title = {Building New Structural Distortion Descriptors through Pressure Engineering toward Enhanced Violet Emission in 2D Hybrid Perovskite},
journal = {Advanced Optical Materials},
volume = {12},
number = {35},
pages = {2401732},
doi = {https://doi.org/10.1002/adom.202401732},
year = {2024}
}

@article{50,
author = {Yu, Xihan and Fang, Yuanyuan and Sun, Xuening and Xie, Ying and Liu, Cailong and Wang, Kai and Xiao, Guanjun and Zou, Bo},
title = {Pressure-Tuning Localized Excitons Toward Enhanced Emission, Photocurrent Enhancement and Piezochromism in Unconventional ACI-Type 2D Hybrid Perovskites},
journal = {Angewandte Chemie International Edition},
volume = {63},
number = {46},
pages = {e202412756},
doi = {https://doi.org/10.1002/anie.202412756},
year = {2024}
}

@article{51,
author = {Ma, Zhiwei and Lv, Pengfei and He, Xin and Wang, Feng and Li, Yongguang and Xiao, Guanjun and Zou, Bo},
title = {Self-Trapped Excitons or Bi3+ Ions for Broad Emission in a Lead-Free Double Perovskite? Hearing What Pressure Says},
journal = {Nano Letters},
volume = {25},
number = {23},
pages = {9345-9352},
year = {2025},
doi = {10.1021/acs.nanolett.5c01709},
}

@Article{52,
author ="Hazra, Vishwadeepa and Mandal, Arnab and Bhattacharyya, Sayan",
title  ="Optoelectronic insights of lead-free layered halide perovskites",
journal  ="Chem. Sci.",
year  ="2024",
volume  ="15",
issue  ="20",
pages  ="7374-7393",
publisher  ="The Royal Society of Chemistry",
doi  ="10.1039/D4SC01429D",
}

@Article{53,
author ="Romani, Lidia and Bala, Anu and Kumar, Vijay and Speltini, Andrea and Milella, Antonella and Fracassi, Francesco and Listorti, Andrea and Profumo, Antonella and Malavasi, Lorenzo",
title  ="PEA2SnBr4: a water-stable lead-free two-dimensional perovskite and demonstration of its use as a co-catalyst in hydrogen photogeneration and organic-dye degradation",
journal  ="J. Mater. Chem. C",
year  ="2020",
volume  ="8",
issue  ="27",
pages  ="9189-9194",
publisher  ="The Royal Society of Chemistry",
doi  ="10.1039/D0TC02525A",
}

@article{54,
author = {Straus, Daniel
B. and Kagan, Cherie R.},
title = {Electrons, Excitons, and Phonons in Two-Dimensional Hybrid Perovskites: Connecting Structural, Optical, and Electronic Properties},
journal = {The Journal of Physical Chemistry Letters},
volume = {9},
number = {6},
pages = {1434-1447},
year = {2018},
doi = {10.1021/acs.jpclett.8b00201},
}

@article{55,
author = {Jin, Linrui and Mora Perez, Carlos and Gao, Yao and Ma, Ke and Park, Jee Yung and Li, Shunran and Guo, Peijun and Dou, Letian and Prezhdo, Oleg and Huang, Libai},
title = {Superior Phonon-Limited Exciton Mobility in Lead-Free Two-Dimensional Perovskites},
journal = {Nano Letters},
volume = {24},
number = {12},
pages = {3638-3646},
year = {2024},
doi = {10.1021/acs.nanolett.3c04895},
}

@article{56,
author = {Srimath Kandada, Ajay Ram and Silva, Carlos},
title = {Exciton Polarons in Two-Dimensional Hybrid Metal-Halide Perovskites},
journal = {The Journal of Physical Chemistry Letters},
volume = {11},
number = {9},
pages = {3173-3184},
year = {2020},
doi = {10.1021/acs.jpclett.9b02342},
}

@article{57,
author    = {Zhou, Hongzhi and Feng, Qingjie and Sun, Cheng and Li, Yahui and
               Tao, Weijian and Tang, Wei and Li, Linjun and Shi, Enzheng and
               Nan, Guangjun and Zhu, Haiming},
title     = {Robust excitonic light emission in 2D tin halide perovskites by weak excited state polaronic effect},
journal   = {Nature Communications},
year      = {2024},
volume    = {15},
number    = {1},
pages     = {8541},
doi       = {10.1038/s41467-024-52952-9},
}

@article{58,
author = {Tao, Weijian and Zhang, Yao and Zhu, Haiming},
title = {Dynamic Exciton Polaron in Two-Dimensional Lead Halide Perovskites and Implications for Optoelectronic Applications},
journal = {Accounts of Chemical Research},
volume = {55},
number = {3},
pages = {345-353},
year = {2022},
doi = {10.1021/acs.accounts.1c00626},
}

@Article{59,
author ="Dybała, Filip and Kudrawiec, Robert and Polak, Maciej P. and Herman, Artur P. and Sieradzki, Adam and Mączka, Mirosław",
title  ="Near-bandgap emission in [HOC2H4NH3]2PbI4 perovskite under hydrostatic pressure: emission of a free exciton and a polaronic exciton",
journal  ="Mater. Adv.",
year  ="2025",
volume  ="6",
issue  ="2",
pages  ="569-578",
publisher  ="RSC",
doi  ="10.1039/D4MA00743C",
}

@article{60,
title = {Free and Self-Trapped Exciton Emission in Perovskite CsPbBr$_3$ Microcrystals},
author = {Pan, Fang and Li, Jinrui and Ma, Xiaoman and Nie, Yang and Liu, Beichen and Ye, Honggang},
journal = {RSC Adv.},
year = {2022},
volume = {12},
pages = {1035-1042},
doi = {10.1039/D1RA08629D},
}

\end{document}